%% file: ms.tex
\documentclass[10pt,a4paper,oneside]{article}

\usepackage[english]{babel}
\usepackage[utf8]{inputenc}
\usepackage[T1]{fontenc}

\usepackage{lmodern}

\usepackage{amsmath}
\usepackage{amssymb}
\usepackage{amsthm}

\usepackage{geometry}

\usepackage{authblk}

\usepackage[runin]{abstract}

\usepackage{titlesec}

\titleformat{\section}[block]{\large\bfseries}{\thesection}{1.0em}{}
\titleformat{\subsection}[runin]{\normalsize\bfseries}{\thesubsection}{0.5em}{}
\titleformat{\paragraph}[runin]{\normalsize\itshape}{\theparagraph}{}{}

\usepackage[section]{tocbibind}

\usepackage{pst-tree}

\input{com}

\usepackage[unicode,bookmarksnumbered,pagebackref]{hyperref} 

\hypersetup{
 pdftitle={Extraction of Efficient Programs in IΣ₁-arithmetic},
 pdfauthor={Ján Komara and Paul J. Voda},
 pdfsubject={Logic in Computer Science},
 pdfkeywords={
  IΣ₁-arithmetic,
  Primitive recursive functions,
  Program extraction,
  Induction
 },
}

\hypersetup{
 unicode,
 colorlinks=true,
 allcolors=[rgb]{0,0.5,0.5},
 urlcolor=[rgb]{0.58,0.0,0.83},
 breaklinks=true,
}

\begin{document}


\title{Extraction of Efficient Programs\\ in \texorpdfstring{\ISA}{IΣ₁}-arithmetic}

\author{Ján Komara}
\author{Paul J. Voda}

\affil{
Institute of Informatics, Faculty of Mathematics and Physics,
\\
Comenius University, Mlynská dolina, 842 15 Bratislava, Slovakia.
\\
Technical Report July, 2000
}

\date{}

\maketitle


\renewcommand{\abstractname}{Abstract.}

\begin{abstract}
{\em C\/}lausal {\em L\/}anguage (CL) is a declarative programming and
verifying system used in our teaching of computer science. CL
is an implementation of, what we call, $\PR{+}\ISA$ paradigm (primitive
recursive functions with \ISA-arithmetic). This paper introduces an extension
of \ISA-proofs called {\em extraction\/} proofs where one can extract from the
proofs of $\Pi_2$-specifications primitive recursive programs as efficient
as the hand-coded ones. This is achieved by having the programming
constructs correspond exactly to the proof rules with the computational
content.
\end{abstract}


\input{sc1}    
\input{sc2}    
\input{sc3}    
\input{sc4}    


\input{ms.bbl} 

\end{document}

%% file: com.tex
\newcommand{\SECT}[1]{\section{#1}}
\newcommand{\PAR}[1]{\subsection{{#1\texorpdfstring{.}{}}}}

\numberwithin{equation}{subsection}
\renewcommand{\theequation}{\arabic{equation}}

\newcommand{\XS}[1]{Sect.~\ref{sc:#1}}
\newcommand{\XP}[1]{Par.~\ref{par:#1}}

\newcommand{\XEL}[2]{(\ref{eq:#1\if #2\@\else:#2\fi})}
\newcommand{\XEG}[2]{\ref{par:#1}(\ref{eq:#1\if #2\@\else:#2\fi})}

\DeclareMathAlphabet{\bmit}{OT1}{cmr}{bx}{it}

\newenvironment{CLAUSES}{%
\it\small\begin{tabbing}%
=\==\==\====================\=======\=\kill}{%
\end{tabbing}}

\newcommand{\pstreeJKa}{
 \pstree[edge=none,
         levelsep=*0.1cm
        ]}
\newcommand{\pstreeJKb}{
 \pstree[edge={\ncline[linestyle=dotted]},
         treesep=1.5cm,
         treefit=loose,
         levelsep=*0.1cm
        ]}
\newcommand{\pstreeJKai}[2]{%
\if #2 {#1} \else \pstreeJKa{#1}{#2} \fi}

\newcommand{\TrJK}[2]{\Tr[#1]{\ensuremath{#2}}}
\newcommand{\TrJKl}[3]{\Tr[#1]{\ensuremath{#2}}~[tnpos=r,tnsep=2pt]{#3}}
\newcommand{\pstreeJKbb}{
 \pstree[edge={\ncline[linestyle=dotted]},
         treesep=1.5cm,
         treefit=loose,
         levelsep=*0.2cm
        ]}
\newcommand{\pstreeJKc}{
 \pstree[edge={\ncline[linestyle=dotted]},
         levelsep=*0.3cm
        ]}

\newcounter{psequation}
\newenvironment{pslabel}{%
 \refstepcounter{equation}%
 (\theequation)%
 }{}
\newcommand{\LABEL}[1]{\begin{pslabel}\label{#1} \end{pslabel}}

\newcommand{\IT}[1]{{\it {#1}\/}}
\newcommand{\ISA}{\ensuremath{I\Sigma_1}}
\newcommand{\PR}{\ensuremath{\IT{PR}}}
\newcommand{\EQx}[2]{\ensuremath{\stackrel{\text{#1}}{#2}}}
\newcommand{\mLRW}{\ensuremath{\Leftrightarrow}}
\newcommand{\RW}{\rightarrow}
\newcommand{\LW}{\leftarrow}
\newcommand{\LRW}{\leftrightarrow}
\newcommand{\ARR}[1]{\ensuremath{\Vec{#1}}}   
\newcommand{\ARRi}[2]{\ensuremath{\Vec{#1}_{#2}}}   
\newcommand{\GOAL}{\ensuremath{\ast}}   
\newcommand{\SETl}{\ensuremath{\mathrel{:=}}}
\newcommand{\SETr}{\ensuremath{\mathrel{=:}}}
\newcommand{\NUM}[1]{\ensuremath{#1}}   
\newcommand{\N}{\ensuremath{\mathbb{N}}}
\newcommand{\RED}{\blacktriangleright}
\newcommand{\GR}{\asymp}         
\def\IN{\ensuremath{\mathrel{\varepsilon}}}

\newcommand{\oPRD}{\ensuremath{\mathop{\hbox{\rlap{\kern2.5pt\raise3.5pt\hbox{.}}$-$}}}}
\newcommand{\oTIMES}{\ensuremath{{\cdot}}}
\newcommand{\aSQRT}[1]{\ensuremath{[ \sqrt{#1} ]}}
\newcommand{\oDIVp}{\ensuremath{\mathop{\div_1}}}

\newcommand{\aS}[2]{{\ensuremath{{#1}\text{\rm\small\bf {#2}}}}}

\newcommand{\aSZP}[1]{\ensuremath{|{#1}|}}

\newcommand{\nRT}{\ensuremath{\IT{Flat}}}

\newcommand{\nBT}{\ensuremath{\IT{Bt}}}
\newcommand{\nND}{\ensuremath{\IT{Nd}}}
\newcommand{\aSZBT}[1]{\ensuremath{|{#1}|_b}}
\newcommand{\rINBT}{\ensuremath{\mathrel{\in_b}}}

\newcommand{\nBST}{\ensuremath{\IT{Bst}}}
\newcommand{\aLTBST}[2]{\ensuremath{{#1} \prec {#2}}}
\newcommand{\aGTBST}[2]{\ensuremath{{#1} \succ {#2}}}
\newcommand{\rINBST}{\ensuremath{\mathrel{\in_s}}}
\newcommand{\aINS}[2]{\ensuremath{\{ {#1} \} \cup {#2}}}

\newcommand{\MYLN}[1]{\if #1 0mm \else #1\baselineskip \fi}

\newcommand{\MYPS}[3]{%
 \begin{center}\small%
 {#3}%
 \end{center}}
\newcommand{\MYPSa}[3]{%
 \vspace{\MYLN{#1}}%
 \begin{center}\small%
 {#3}%
 \end{center}%
 \vspace{\MYLN{#2}}}

%% file: sc1.tex
\SECT{Introduction}\label{sc:intro}

The class of effectively computable functions over natural numbers coincides
by the thesis of Church with recursive functions as defined by
Herbrand-G\"odel style equations. We use Herbrand-G\"odel-like recursive
equations because they offer the programming comfort with almost
unrestricted kinds of recursion and the computation of recursive
equations by reductions permits a fine degree of control over the length of
reduction sequences. We interpret the recursive equations into natural
numbers because the concept of truth in \N\ is well understood even by
beginners and the theory of recursive functions and arithmetic offers a firm
natural semantical background.

A possible objection by computer scientists that the domain \N\ means
unpleasant coding (arithmetization) of the rich set of data structures used
in
programming is answered by coding into \N\ in the style of LISP with a
pairing function (instead of \IT{cons}). We obtain a degree of comfort as it
is known from declarative
prog.
languages (say Haskell). The examples in \XS{ex} should convince the reader.
An objection that the coding may prolong the length of reduction sequences
is answered by computing in mixed representation (see \XS{ari}).

We are interested not only in a programming language but also in the
verification of its programs. By restricting ourselves to \N\ we can use the
simplest of formal theories: Peano arithmetic. Since the computationally
feasible functions are a proper subset of elementary functions and the
latter are but a tiny subset of primitive recursive functions (\PR), we
restrict the strength of our system to \ISA-arithmetic where the induction
formulas are $\Sigma_1$ (see \cite{qq:HajekPudlak:93}).  The provably
recursive functions of \ISA\ are exactly the primitive recursive functions. We can thus
call our approach the {\em $\PR{+}\ISA$ programming/verifying paradigm}.  
We will
briefly discuss our computer implementation of the paradigm in the
form of the system CL
({\em C\/}lausal {\em L\/}anguage)
in the conclusion of this paper.

In order to prove properties of functions defined by a rich variety of
recursion schemes we need a rich variety of induction schemes. The schemes
are needed for the programming/verifying comfort but their power
does not exceed
\PR\ functions and
mathematical induction with $\Sigma_1$-formulas.
Our results are of
metamathematical
(proof-theoretical)
character because we do a rigorous
development of a usable programming language (usable at least in the
teaching if not yet in real life) within the theory of programming
languages. This means in our case
that we investigate how to express programs as
primitive recursive functions and how to do the proofs of their properties
in \ISA-arithmetic.

Beside the \ISA-characterization of our language in \XS{char},
the main contribution of this paper is the introduction of a special kind of
{\em extraction proofs\/} 
from which efficient programs
for functions satisfying $\Pi_2$-specifications can be extracted. The main
idea is that the user,
by deciding which rules have computational content
and which not, 
can control the efficiency of the extracted witnessing
function exactly as if he first hand-coded it and then proved that
it satisfies the specifications. The goal is achieved by having the programming
constructs correspond exactly to the proof rules with the computational
content. This contrasts with the approach in systems such as
PX \cite{qq:PX:88}
and MINLOG \cite{qq:MINLOG:98} where the decision which rules should have
computational content is automatic.

%% file: sc2.tex
\SECT{Primitive Recursive Functions and \texorpdfstring{\ISA}{IΣ₁}-arithmetic}\label{sc:ari}

We now give a brief overview of recursion-theoretic semantics so the
reader can see how the proofs of $\Pi_2$-specifications which will be
discussed in \XS{ex} can control the efficiency of extracted programs (for
details see
\cite{qq:Voda:94:CSL94,qq:KomaraVoda:98:CSL98,qq:Voda:theory:00:www}).
Primitive recursive functions are discussed in detail for instance
in \cite{qq:Peter:67,qq:Rose:84}.

\PAR{Mixed representation of natural numbers}

Every positive number $x$ can be uniquely written in a form
\(
x  = \sum_{i < n} d_i \oTIMES 2^i
\)
as a sequence
\(
d_{n-1}d_{n-2}\ldots d_1d_0
\)
of {\em dyadic digits\/} $1 \leq d_i \leq 2$ and we have
\(
n = \Theta(\log(x))
\).
Every natural number $x$ can be obtained in the {\em dyadic
representation\/} by finitely many applications to the constant $0$
of {\em dyadic successor\/}
functions
\(
\aS{x}{1} = 2\oTIMES x{+}1
\)
and 
\(
\aS{x}{2} = 2\oTIMES x{+}2
\):
\(
\aS{\aS{\aS{0}{d$_{n-1} \cdots$}}{d$_1$}}{d$_0$}
\).

For a suitable {\em pairing\/} function $(x,y)$ we have
{\small
\begin{gather*}
(x_1,y_1) = (x_2,y_2) \RW x_1 = x_2 \wedge y_1 = y_2
\\
x < (x,y) \wedge y < (x,y)
\\
x = 0 \vee \exists v \exists w\, x = (v,w)
\ .
\end{gather*}
} \!\!\!\!\!
We require that for the {\em pair size\/} function $\aSZP{x}$ satisfying
\(
\aSZP{0} = 0
\)
and
\(
\aSZP{(x,y)} = \aSZP{x}+\aSZP{y}+1
\)
we have
\(
\aSZP{x} = \Theta(\log(x))
\).
Note that from the second property we get $0 \neq (x,y)$ and so every
natural number can be uniquely obtained in the {\em pair representation\/}
by finitely many applications of $(x,y)$ to the constant $0$. Thus \N\ can
be identified with the S-expressions of LISP (with the single atom $\IT{nil}
= 0$). When there can be no confusion of pairing with the comma separating
arguments of functions we abbreviate $(x,y)$ to $x,y$. We also write $x,y,z$
for $x,(y,z)$. To every finite sequence of numbers
\(
x_1 \dotsc x_n
\)
there is exactly one natural number, namely
\(
x_1, \dotsc, x_n, 0
\),
coding the sequence as a {\em list\/}.

{\em Mixed numerals\/} are terms obtained from $0$ by dyadic successors and
pairing. This {\em mixed representation\/} of \N\ is not unique but permits
the mixed mode computation of arithmetic and symbolic operations without unnecessary
conversions between dyadic and pair representations (provided the definitions are
well-typed).

\PAR{Recursive definitions}

We need three variable binding term operators, respectively called {\em
let\/}, {\em dyadic discrimination\/}, and {\em pair discrimination\/}
terms.  The terms bind the indicated variables in the indicated positions:
{\small
\begin{align*}
L_y(x,\tau[y]) = z & \LRW \exists y (x = y \wedge z = \tau[y])
\\
D_v(x,y,\tau_1[v],\tau_2[v]) = z & \LRW {}
 x = 0 \wedge z = y \vee 
\\
& \phantom{{} \LRW {}}
\exists v( x = \aS{v}{1} \wedge z = \tau_1[v]) \vee 
     \exists v( x = \aS{v}{2} \wedge z = \tau_2[v])
\\
P_{v,w}(x,y,\tau[v,w]) = z & \LRW {} x = 0 \wedge z = y \vee
 \exists v \exists w ( x = v,w \wedge z = \tau[v,w] )
\ .
\end{align*}
} \!\!\!\!
{\em Recursive\/} terms are constructed from variables and $0$ by the three
operators and by applications of functions
$\aS{x}{1}$, $\aS{x}{2}$, $(x,y)$, 
as well as of {\em defined\/} partial functions. For every recursive term
$\tau[f;\ARR{x}]$ we let the function symbol $\lambda \ARR{x}.\tau$ to denote the
least partial function $f$ solving the functional equation
\(
f(\ARR{x}) \simeq \tau[f;\ARR{x}]
\).
The equation is a {\em recursive definition\/} of $f$.  Note that
\(
\lambda \ARR{x}.\tau
\)
is not the standard lambda notation because the (meaningless) {\em
recursive\/} function symbol $f$ can be applied in $\tau$. The function
symbols
\(
\lambda \ARR{x}.\tau
\)
bind the variables $\ARR{x}$ as well as applications of $f$. A recursive
term is {\em closed\/} if it contains no free variables and no free
applications of the recursive symbol $f$.

The partial functions definable by recursive definitions are exactly the
partial recursive functions. The style of the definitions is basically that
of Herbrand-G\"odel recursive equations. We present the recursive
definitions in \XS{ex} in the form of {\em clausal definitions\/} which are
only {\em unfolded\/} recursive definitions. Clausal definitions are used to
increase the readability.

\PAR{Computation by reductions}

The reason for our recursive terms is that one can obtain as efficient
reduction (computation) sequences as one wishes. We reduce a closed
recursive term $\tau$ to a mixed numeral $\rho$, in writing
\(
\tau \RED \rho
\),
by repeatedly locating in it the leftmost {\em redex\/} and rewriting it by
its {\em contractum\/}. {\em Let\/} redexes (and contracta) are
\(
L_y(\rho,\tau[y]) \RED \tau[\rho]
\),
{\em pair\/} redexes are
\(
P_{v,w}(0,\tau_1,\tau_2[v,w]) \RED \tau_1
\)
and
\(
P_{v,w}(\rho,\tau_1,\tau_2[v,w]) \RED \tau_2[\rho_1,\rho_2]
\)
where $\rho$ is a mixed numeral and in the last redex we have
\(
\rho = \rho_1,\rho_2
\).
Note that this may mean a conversion if the outermost application of $\rho$
is not pairing. The {\em dyadic\/} redexes involving $D_v$ are similar. {\em
Lambda\/} redexes are
\(
(\lambda \ARR{x}.\tau[f;\ARR{x}])(\ARR{\rho}) \RED 
 \tau[\lambda \ARR{x}.\tau;\ARR{\rho}]
\)
where $\ARR{\rho}$ is an $n$-tuple of mixed numerals. Note that the mixed
numerals are irreducible.

The denotational semantics is given by recursive definitions and the
operational one by reductions. Both semantics coincide because for a partial
function
\(
f = \lambda \ARR{x}.\tau
\)
we have $f(\ARR{x}) \simeq y$ iff
\(
(\lambda \ARR{x}.\tau)(\ARR{\rho}) \RED \rho_0
\)
for some mixed numerals $\ARR{\rho}$ denoting the corresponding numbers
$\ARR{x}$ and $\rho_0$ denoting $y$. Note that the
function symbol
\(
\lambda \ARR{x}.\tau
\)
extensionally  denotes a partial function $f$ while intensionally it is a
{\em program\/} (algorithm) for the computation of $f$.

\PAR{\texorpdfstring{\ISA}{IΣ₁}-arithmetic
\texorpdfstring{{\rm \cite{qq:HajekPudlak:93}}}{[HP93]}}

Our proof system for \ISA-arithmetic is based on positive, i.e.
non-refutational, tableaux (see
\cite{qq:Smullyan:68,qq:KomaraVoda:98:CSL98}).
We mark in the tableaux shown in \XS{ex} the {\em goal\/} formulas as $\phi
\GOAL$ and leave the {\em assumption\/} formulas unmarked. We work in
recursive extensions of \ISA\
(see \cite{qq:Shoenfield:67});
also denoted by the same symbol \ISA. We
have also a rich set of admissible rules (see \XS{ex}) for the proofs of
properties of recursively defined functions.

The {\em graph\/} $f(\ARR{x}) \GR y$ of every partial recursive function 
\(
f = \lambda \ARR{x}.\tau
\)
is $\Sigma_1$-definable in \ISA\ and we have
\(
\ISA \vdash f(\ARR{x}) \simeq \tau[f;\ARR{x}]
\).
The last means
\(
\ISA \vdash f(\ARR{x}) \GR y \LRW \tau[f;\ARR{x}] \GR y
\)
where $\tau \GR y$ stands for the graph of the partially denoting term
$\tau$.  Graphs of terms are defined in the obvious way such that for every
closed recursive term $\tau$ we have: $\tau \GR y$ iff $\tau \RED
\rho$ for a mixed numeral $\rho$ denoting $y$.

A recursive function
\(
f = \lambda \ARR{x}.\tau
\)
is {\em provably recursive in \ISA\ (total, terminating)\/} if
\(
\ISA \vdash \exists y\,f(\ARR{x}) \GR y
\). 
We then have
\(
\ISA \vdash f(\ARR{x}) = \tau[f;\ARR{x}]
\)
and the graph of $f$ is also $\Pi_1$-definable in \ISA.  It is well-known
\cite{qq:Kreisel:52}) that the provably recursive functions in \ISA\ are
exactly the primitive recursive functions.

\PAR{Satisfying \texorpdfstring{$\Pi_2$}{Π₂}-specifications}\label{par:specs}

We wish to find a {\em witness\/} function $f(\ARR{x})$ for a {\em
$\Pi_2$-specification\/} sentence
\(
\forall \ARR{x}\,(\psi[\ARR{x}] \RW \exists y \phi[\ARR{x},y])
\)
which is possibly under assumption $\psi[\ARR{x}]$.  Instead of proving the
specification sentence we will prove by an {\em extraction\/} proof its {\em
witnessing\/} formula:
\begin{align}
\ISA \vdash_e
\psi[\ARR{x}] \RW \phi[\ARR{x},f(\ARR{x})]\ .
\label{eq:specs:witness}
\end{align}
Here $\vdash_e$ stands for {\em extraction provability\/} and the symbol
$f(\ARR{x})$ should be understood as an `unknown' which obtains a value in
the extraction proof by a {\em definition\/}
\(
f(\ARR{x}) \SETl \tau
\).
It will be clear from the examples in \XS{ex} that given an extraction proof
of \XEL{specs}{witness} we can primitively recursively find a term $\tau$
and a (standard) proof of
\begin{align}
\ISA + \forall \ARR{x}\, f(\ARR{x}) \simeq \tau \vdash 
\psi[\ARR{x}] \RW
\exists y( \phi[\ARR{x},y] \wedge f(\ARR{x}) \GR y )
\ .
\label{eq:specs:trans}
\end{align}
The proof of \XEL{specs}{trans} decomposes into proofs of the {\em partial
correctness\/} formula
\(
\psi[\ARR{x}] \wedge f(\ARR{x}) \GR y \RW \phi[\ARR{x},y]
\)
and of the {\em termination\/} formula
\(
\psi[\ARR{x}] \RW \exists y\,f(\ARR{x}) \GR y
\).
That partial functions are needed in order to obtain efficiency will be seen
in \XP{ex:mind}.

\PAR{Computational content of extraction proof rules}

We can use in extraction proofs four kinds of proof rules involving the
unknowns which correspond exactly to the kinds of programming constructs
allowed in recursive definitions. By suitable applications of rules in an
extraction proof of the witnessing formula we can guide the construction of
the function to be extracted in exactly the same way as if we did the
definition by hand.  The rules are:
\begin{center}
\begin{tabular}{|c|c|}
\hline
\hline
extraction rule: &
corresponding standard proof rule:
\\
\hline
recursion &
induction
\\
discrimination &
cut
\\
assignment $\tau \SETr x$ &
use of theorem $\exists x\, \tau = x$
\\
\ definition $f(\ARR{\rho}) \SETl \rho$\ {} &
\ consequence of the defining axiom $f(\ARR{x}) \simeq \tau$\ {}
\\
\hline
\hline
\end{tabular}   
\end{center}
Standard tableau rules are permitted in a branch above an application of a
definition rule only if they are devoid of {\em computational content\/}
affecting termination of $f$. Below a definition rule (when a clause of a
definition was completed) the branch can closed by unrestricted tableau
rules. Rules without computational content affecting termination include the
rule of replacement of equals (Leibnitz rule), term simplification rules by
using open \ISA-theorems, propositional simplification rules, and quantifier
rules (see \XP{ex:mind}). Such rules may introduce no new variables into the
definition being constructed.

Standard propositional tableau rules have in general computational content
affecting termination and can be used in branches above definition rules
only when the $\vdash_e$-proof checker is able to determine that they are
used safely, i.e. without affecting termination.  In particular, all
non-commented uses of {\em assumption\/} rules
\(
(\phi_1 \RW \phi_2)\!\GOAL / \phi_1, \phi_2 \GOAL
\)
given in the examples of \XS{ex} are safe. 

%% file: sc3.tex
\SECT{Examples of Extraction of Programs in \texorpdfstring{\ISA}{IΣ₁}-arithmetic}\label{sc:ex}

\PAR{Discrimination on predicates}\label{par:ex:case}

Assume that \ISA\ proves that for every \ARR{x} exactly one of
\(
R_1(\ARR{x}), \dotsc, R_n(\ARR{x})
\)
holds. We can then use the {\em discrimination\/} (case analysis) rule as
shown in the following on the left:
\MYPSa{-0.3}{-0.2}{
\pstree[edge={\ncline[linestyle=dotted]},
         treesep=0.5cm,
         treefit=loose,
         levelsep=*0.3cm
        ]{
 \Tr{}
 }{
 \pstreeJKai{
  \TrJK{}{R_1(\ARR{x})}
  }{
  }
 \pstreeJKai{
  \TrJK{}{R_2(\ARR{x})}
  }{
  }
 \pstreeJKai{
  \TrJK{edge=none}{\dotsb}
  }{
  }
 \pstreeJKai{
  \TrJK{}{R_n(\ARR{x})}
  }{
  }
 }
\pstreeJKa{
 \TrJKl{}{\phi[\ARR{x},f(\ARR{x})]}{\GOAL}
 }{
 \pstree[edge={\ncline[linestyle=dotted]},
         treesep=0.2cm,
         treefit=loose,
         levelsep=*0.3cm
        ]{
  \Tr{}
  }{
  \pstreeJKc{
   \TrJK{}{R_1(\ARR{x})}
   }{
   \pstreeJKai{
    \TrJK{}{f(\ARR{x}) \SETl \tau_1[\ARR{x}]}
    }{
    }
   }
  \pstreeJKc{
   \TrJK{}{R_2(\ARR{x})}
   }{
   \pstreeJKai{
    \TrJK{}{f(\ARR{x}) \SETl \tau_2[\ARR{x}]}
    }{
    }
   }
  \pstreeJKai{
   \TrJK{edge=none}{\dotsb}
   }{
   }
  \pstreeJKc{
   \TrJK{}{R_n(\ARR{x})}
   }{
   \pstreeJKai{
    \TrJK{}{f(\ARR{x}) \SETl \tau_n[\ARR{x}]}
    }{
    }
   }
  }
 }
}
The rule is applied on the right. The extracted witnessing
function $f$ has then the following definition:
\begin{CLAUSES}
\> $f(\ARR{x}) = \tau_1[\ARR{x}] \LW R_1(\ARR{x})$
\\
\> $f(\ARR{x}) = \tau_2[\ARR{x}] \LW R_2(\ARR{x})$
\\
\> $\dotsm$
\\
\> $f(\ARR{x}) = \tau_n[\ARR{x}] \LW R_n(\ARR{x})$
\ .
\end{CLAUSES}
Consider the minimum function $\min(x,y)$ with the specification:
\(
\forall x \forall y \exists m
\bigl(m \leq x \wedge m \leq y \wedge (m = x \vee m = y)\bigr)
\).
The proof of the witnessing formula \XEL{ex:case}{min:spec:proof} starts
with the {\em dichotomy\/} discrimination on $x \leq y$ and $x > y$ where in
each of the branches the obvious definition rules for $\min(x,y)$ close the
tableau:
\MYPSa{-0.3}{-0.2}{
\pstreeJKa{
 \TrJKl{}{\min(x,y) \leq x \wedge \min(x,y) \leq y \wedge 
         (\min(x,y) = x \vee \min(x,y) = y)}{\GOAL\, \LABEL{eq:ex:case:min:spec:proof}}
 }{
 \pstreeJKb{
  \Tr{}
  }{
  \pstreeJKa{
   \TrJK{}{x \leq y}
   }{
   \pstreeJKai{
    \TrJK{}{\min(x,y) \SETl x}
    }{
    }
   }
  \pstreeJKa{
   \TrJK{}{x > y}
   }{
   \pstreeJKai{
    \TrJK{}{\min(x,y) \SETl y}
    }{
    }
   }
  }
 }
}
The extracted definition is:
\begin{CLAUSES}
\> $\min(x,y) = x \LW x \leq y$
\\
\> $\min(x,y) = y \LW x > y$
\ .
\end{CLAUSES}

\PAR{Discrimination on patterns}

Assume that \ISA\ proves that for every $x$
exactly one of 
\(
\exists \ARRi{y}{1}\, z = \rho_1[\ARRi{y}{1}]
\),
\ldots, 
\(
\exists \ARRi{y}{n}\, z = \rho_n[\ARRi{y}{n}]
\)
holds.  Here the terms $\rho_i[\ARRi{y}{i}]$ are {\em patterns\/} and must be
such that when $z = \rho_i[\ARRi{y}{i}]$ holds then the numbers
$\ARRi{y}{i}$ are unique and can be primitively recursively obtained from
$z$. For every term $\tau$ we can then use a rule {\em discriminating on
patterns\/}:
\MYPS{-0.3}{-0.2}{
\pstree[edge={\ncline[linestyle=dotted]},
         treesep=0.5cm,
         treefit=loose,
         levelsep=*0.3cm
        ]{
 \Tr{}
 }{
 \pstreeJKai{
  \TrJK{}{\tau = \rho_1[\ARRi{y}{1}]}
  }{
  }
 \pstreeJKai{
  \TrJK{}{\tau = \rho_2[\ARRi{y}{2}]}
  }{
  }
 \pstreeJKai{
  \TrJK{edge=none}{\dotsb}
  }{
  }
 \pstreeJKai{
  \TrJK{}{\tau = \rho_n[\ARRi{y}{n}]}
  }{
  }
 }
}
where the variables $\ARRi{y}{i}$ are new, i.e. {\em eigenvariables\/}.

For instance, the $0$-$1$ valued {\em signum\/} function $\IT{sg}(x)$ has
the specification:
\(
\forall x \exists s
(x = 0 \wedge s = 0 \vee x > 0 \wedge s = 1)
\)
and it is extracted with the clauses
$\IT{sg}(0) = 0$
and
$\IT{sg}(y+1) = 1$
from the following proof:
\MYPS{-0.3}{-0.2}{
\pstreeJKa{
 \TrJKl{}{x= 0 \wedge \IT{sg}(x) = 0 \vee x > 0 \wedge \IT{sg}(x) = 1}{\GOAL}
 }{
 \pstree[edge={\ncline[linestyle=dotted]},
         treesep=0.2cm, 
         treefit=loose, 
         levelsep=*0.2cm
        ]{
  \Tr{}
  }{
  \pstreeJKa{
   \TrJK{}{x = 0}
   }{
   \pstreeJKai{
    \TrJK{}{\IT{sg}(x) \SETl 0}
    }{
    }
   }
  \pstreeJKa{
   \TrJK{}{x = y+1}
   }{
   \pstreeJKai{
    \TrJK{}{\IT{sg}(x) \SETl 1}
    }{
    }
   }
  }
 }
}
by a rule discriminating on patterns $x = 0$ and $\exists y\, x = y+1$.

\PAR{Monadic representation of natural numbers}\label{par:ex:n}

Every natural number is obtained in {\em monadic\/} representation by
finitely many applications of the (monadic) {\em successor\/} function $x{+}1$
to the constant $0$:
\(
(\dotsm ((0{+}1){+}1) \dotsm) {+}1
\).
Monadic representation is used in the rules of {\em induction on monadic
notation\/},
better known as {\em mathematical induction\/} shown
on the left
($\phi$ is $\Sigma_1$):
\MYPS{-0.3}{-0.2}{
\pstreeJKa{
 \TrJKl{}{\phi[x]}{\GOAL}
 }{
 \pstreeJKb{
  \Tr{}
  }{
  \pstreeJKai{
   \TrJKl{}{\phi[0]}{\GOAL}
   }{
   }
  \pstreeJKai{
   \TrJKl{}{\phi[x] \RW \phi[x+1]}{\GOAL}
   }{
   }
  }
 }
\pstreeJKa{
 \TrJKl{}{\phi[x,f(x,\ARR{z}),\ARR{z}]}{\GOAL}
 }{
 \pstree[edge={\ncline[linestyle=dotted]},
         treesep=0.2cm, 
         treefit=loose, 
         levelsep=*0.2cm
        ]{
  \Tr{}
  }{
  \pstreeJKc{
   \TrJKl{}{\phi[0,f(0,\ARR{z}),\ARR{z}]}{\GOAL}
   }{
   \pstreeJKai{
    \TrJK{}{f(0,\ARR{z}) \SETl \tau_1[\ARR{z}]}
    }{
    }
   }
  \pstreeJKa{
   \TrJK{}{\phi[x,f(x,\ARR{z}),\ARR{z}]}
   }{
   \pstreeJKc{
    \TrJKl{}{\phi[x+1,f(x+1,\ARR{z}),\ARR{z}]}{\GOAL}
    }{
    \pstreeJKai{
     \TrJK{}{f(x+1,\ARR{z}) \SETl \tau_2[f(x,\ARR{z}),x,\ARR{z}]}
     }{
     }
    }
   }
  }
 }
}
The corresponding rule of {\em recursion on monadic notation\/} is applied
on the right to
extract a witness $f(x,\ARR{z})$ for the specification
\(
\exists y\, \phi[x,y,\ARR{z}]
\).
Note that in the inductive case we effectively use $f(x,\ARR{z})$
as the eigenvariable for the IH formula
$\exists y\, \phi[x,y,\ARR{z}]$.
The `eigenvariable' is then used in
the definition of $f(x+1,\ARR{z})$. The extracted definition is by
{\em primitive recursion\/}:
\begin{CLAUSES}
\> $f(0,\ARR{z}) = \tau_1[\ARR{z}]$
\\
\> $f(x+1,\ARR{z}) = \tau_2[f(x,\ARR{z}),x,\ARR{z}]$
\ .
\end{CLAUSES}

\PAR{Square root function}\label{par:ex:n:sqrt}

The square root function $\aSQRT{x}$ satisfies the specification:
\(
\forall x \exists y\, y^2 \leq x < (y+1)^2
\).
Its definition: 
\begin{CLAUSES}
\> $\aSQRT{0} = 0$
\\
\> $\aSQRT{x+1} = \aSQRT{x} \LW x+1 < (\aSQRT{x}+1)^2$
\\
\> $\aSQRT{x+1} = \aSQRT{x}+1 \LW x+1 = (\aSQRT{x}+1)^2$
\end{CLAUSES}
is extracted from the following proof by induction on $x$ of its witnessing
formula. In the base case it suffices to define $\aSQRT{0} \SETl 0$ and the
inductive case is:
\MYPSa{-0.1}{-0.1}{
\pstreeJKa{
 \TrJKl{}{\aSQRT{x}^2 \leq x < (\aSQRT{x}+1)^2}{\ \ (IH)}
 }{
 \pstreeJKa{
  \TrJKl{}{\aSQRT{x+1}^2 \leq x+1 < (\aSQRT{x+1}+1)^2}{\GOAL\, 
      \LABEL{eq:ex:n:sqrt:spec:proof}}
  }{
  \pstreeJKb{
  \Tr{}
  }{
   \pstreeJKa{
    \TrJK{}{x+1 < (\aSQRT{x}+1)^2}
    }{
    \pstreeJKai{
     \TrJK{}{\aSQRT{x+1} \SETl \aSQRT{x}}
     }{
     }
    }
   \pstreeJKa{
    \TrJK{}{x+1 = (\aSQRT{x}+1)^2}
    }{
    \pstreeJKai{
     \TrJK{}{\aSQRT{x+1} \SETl \aSQRT{x}+1}
     }{
     }
    }
   \pstreeJKai{
    \TrJK{}{x+1 > (\aSQRT{x}+1)^2}
    }{
    }
   }
  }
 }
}
We use a rule of {\em trichotomy\/} discrimination and in the branch where
\(
x{+}1 < (\aSQRT{x}{+}1)^2
\)
holds we satisfy the goal \XEL{ex:n:sqrt}{spec:proof} by defining
\(
\aSQRT{x{+}1} \SETl \aSQRT{x}
\)
because we have from IH:
\(
\aSQRT{x}^2 \leq x < x{+}1 < (\aSQRT{x}{+}1)^2
\).
When
\(
x{+}1 = (\aSQRT{x}{+}1)^2
\)
holds then the shown definition trivially satisfies
\XEL{ex:n:sqrt}{spec:proof}. The last case
\(
x{+}1 > (\aSQRT{x}{+}1)^2
\)
leads to contradiction with IH, i.e. the case cannot happen, and the
function value is immaterial. Note that the clause corresponding to the
contradictory branch is omitted in the extracted program and yields $0$ by
{\em default\/}.

\PAR{Assignment rules}\label{par:ex:let}

The twofold occurrence of $\aSQRT{x}$ in the recursive clauses in
\XP{ex:n:sqrt} causes the application to be computed twice for each value of
$x$. This leads to the exponential explosion of computation time. The
explosion can be prevented by an {\em assignment\/}: $\aSQRT{x} = z$ folding
to a let term
\(
L_z(\aSQRT{x},\ldots z \ldots)
\):
\begin{CLAUSES}
\> $\aSQRT{0} = 0$
\\
\> $\aSQRT{x+1} = z \LW \aSQRT{x} = z \wedge x+1 < (z+1)^2$
\\
\> $\aSQRT{x+1} = z+1 \LW \aSQRT{x} = z \wedge x+1 = (z+1)^2$
\ .
\end{CLAUSES}
In the following proof we force the assignment to be extracted
by an {\em assignment\/} rule $\aSQRT{x} \SETr z$ with the eigenvariable
$z$:
\MYPSa{-0.3}{-0.2}{
\pstreeJKa{
 \TrJK{}{\aSQRT{x}^2 \leq x < (\aSQRT{x}+1)^2}
 }{
 \pstreeJKa{
  \TrJKl{}{\aSQRT{x+1}^2 \leq x+1 < (\aSQRT{x+1}+1)^2}{\GOAL}
  }{
  \pstreeJKa{
   \TrJK{}{\aSQRT{x} \SETr z}
   }{
   \pstreeJKb{
    \Tr{}
    }{
    \pstreeJKa{
     \TrJK{}{x+1 < (z+1)^2}
     }{
     \pstreeJKai{
      \TrJK{}{\aSQRT{x+1} \SETl z}
      }{
      }
     }
    \pstreeJKa{
     \TrJK{}{x+1 = (z+1)^2}
     }{
     \pstreeJKai{
      \TrJK{}{\aSQRT{x+1} \SETl z+1}
      }{
      }
     }
    \pstreeJKai{
     \TrJK{}{x+1 > (z+1)^2}
     }{
     }
    }
   }
  }
 }
}

\PAR{Measure induction}\label{par:ex:mind}

The rule of {\em measure\/} induction:
\MYPSa{-0.3}{-0.2}{
\pstreeJKa{
 \TrJKl{}{\phi[\ARR{x}]}{\GOAL}
 }{
 \pstreeJKbb{
  \Tr{}
  }{
  \pstreeJKai{
   \TrJKl{}{\forall \ARR{y}(m(\ARR{y}) < m(\ARR{x}) \RW
           \phi[\ARR{y}]) \RW \phi[\ARR{x}]}{\GOAL}
   }{
   }
  }
 }
}
permits to derive the goal $\phi[\ARR{x}]$ under IH that $\phi[\ARR{y}]$
holds for all $\ARR{y}$ having lesser {\em measure\/}:
\(
m(\ARR{y}) < m(\ARR{x})
\).
Note that complete induction on $x$ is a special case of measure induction with
$m(x) = x$.

We can extract a program for the integer division $x \div y$ by the
following proof of the witnessing formula for
\(
y > 0 \RW \exists q\exists r (x = q\oTIMES y+r \wedge r < y)
\)
where we do not treat $y>0$ as an assumption. The proof is by complete
induction on $x$:
\begin{center}\small
\pstreeJKa{
 \TrJKl{}{y > 0 \RW \exists r( x = (x \div  y) \oTIMES y +r \wedge r < y)}
          {\GOAL\ \LABEL{eq:ex:mind:div:goal}}
 }{
 \pstreeJKa{
  \TrJKl{}{\forall x_1 (x_1 < x \RW
          y > 0 \RW
          \exists r (x_1 = (x_1 \div y) \oTIMES y+r \wedge r < y)}{\ (IH)}
  }{
  \pstree[edge={\ncline[linestyle=dotted]},
          treesep=1.0cm,
          treefit=loose,
          levelsep=*0.1cm
         ]{
   \Tr{}
   }{
   \pstreeJKa{
    \TrJK{}{y = 0}
    }{
    \pstreeJKa{
     \TrJKl{}{\top}{\GOAL\ \LABEL{eq:ex:mind:div:goal:1}} 
     }{
     }
    }
   \pstreeJKa{
    \TrJK{}{y \neq 0}
    }{
    \pstreeJKa{
     \TrJKl{}{\exists r(x = (x \div  y) \oTIMES y +r\wedge r < y)}
             {\GOAL\ \LABEL{eq:ex:mind:div:goal:2}}
     }{
     \pstreeJKb{
      \Tr{}
      }{
      \pstreeJKa{
       \TrJK{}{x < y}
       }{
       \pstreeJKai{
        \TrJK{}{x \div y \SETl 0}
        }{
        }
       }
      \pstreeJKa{
       \TrJK{}{x \geq y}
       }{
       \pstreeJKa{
        \TrJKl{}{x = ((x \oPRD y) \div y + 1) \oTIMES y+r \wedge r < y}
                {\ \LABEL{eq:ex:mind:div:inst}}
        }{
        \pstreeJKai{
         \TrJK{}{x \div y \SETl (x \oPRD y) \div y + 1}
         }{
         }
        }
       }
      }
     }
    }
   }
  }
 }
\end{center}
We do first a {\em negation\/} discrimination with $y = 0$. In the case $y =
0$ the goal \XEL{ex:mind}{div:goal} simplifies to
\XEL{ex:mind}{div:goal:1} and so any defining rule for $x {\div} 0$ will do.
In the case $y \neq 0$ we do a dichotomy discrimination. In the case $x < y$
we satisfy the simplified goal \XEL{ex:mind}{div:goal:2} with the
substitution $r \SETl x$ and by defining
\(
x \div y \SETl 0
\).
In the case $x \geq y$ we have $x \oPRD y < x$ and we instantiate IH with
\(
x_1 \SETl x \oPRD y
\)
and use $r$ as eigenvariable whereby we obtain \XEL{ex:mind}{div:inst} after
a simplification. We now satisfy the goal \XEL{ex:mind}{div:goal:2} with the
substitution $r \SETl r$ and with a definition as shown. Instantiations and
existential substitutions have no computational content provided we do not
use in the definition rules eigenvariables (in this case $r$) other than
those coming from assignments and patterns. The extracted definition is
\begin{CLAUSES}
\> $x \div y = 0 \LW y \neq 0 \wedge x < y$
\\
\> $x \div y = (x \oPRD y) \div y + 1 \LW y \neq 0 \wedge x \geq y$
\end{CLAUSES}
where the clause
\(
x \div y = 0 \LW y = 0
\)
is omitted by default. If we decided to apply the assumption rule to
\XEL{ex:mind}{div:goal} without doing the discrimination on $y = 0$ then the
definition extracted from the right branch would not contain the tests $y
\neq 0$ and would still satisfy the partial correctness but not the
termination formula
\(
\exists q\, x \div 0 \GR q
\)
for $y=0$. 

The extracted program is less optimal than it should be due to repeated
tests $y \neq 0$. We obtain a better one when we do the extraction with the
formula:
\(
\exists r (x = (x \oDIVp y) \oTIMES y +r\wedge r < y)
\)
under the assumption $y > 0$. The left branch now disappears and we get a
definition of a partial function $x \oDIVp y$ (diverging when $y = 0$) which
can be explicitly completed to $\div$ as follows:
\begin{CLAUSES}
\> $x \oDIVp y = 0 \LW x < y$
\\
\> $x \oDIVp y = (x \oPRD y) \oDIVp y + 1 \LW x \geq y$
\\
\> $x \div y = x \oDIVp y \LW y > 0$
\ .
\end{CLAUSES}

\PAR{A faster program for \texorpdfstring{$\aSQRT{x}$}{[√x]} with accumulators}\label{par:ex:mind:sqrt}

We can save the squaring operation in the test $x+1 < (z+1)^2$ of \aSQRT{x}
in \XP{ex:let} which is repeatedly done for every recursive call by defining
a function $f(z,a,x)$ with {\em accumulators\/} $z$ and $a$ such that
\(
f(z,z^2,x) = \aSQRT{x}
\)
provided $z^2 \leq x$.
We intend to define $f$ by {\em backward recursion\/} where $a$ grows
towards $x$, i.e. by recursion with the measure $m(z,a,x) = x \oPRD a$.
As $z$ goes to $z{+}1$ the accumulator $a = z^2$ goes to
$a_1 = a{+}2\oTIMES z{+}1 = (z{+}1)^2$. This arrangement reduces the squaring operation
to the increments by $2\oTIMES z {+}1$ which are fast in the dyadic notation.
The definition:
\begin{CLAUSES}
\> $f(z,a,x) = z \LW a+2 \oTIMES z+1 = a_1 \wedge x < a_1$
\\
\> $f(z,a,x) = f(z+1,a_1,x) \LW 
     a+2 \oTIMES z+1 = a_1 \wedge x \geq a_1$
\ .
\end{CLAUSES}
is extracted from the following proof by induction with measure $m$ of
the witnessing formula for 
$z^2 = a \leq x \RW  \exists y\, y^2 \leq x < (y{+}1)^2$:
\MYPS{-0.3}{-0.2}{
\pstreeJKa{
 \TrJKl{}{\forall z_1a_1x_1 (x_1 {\oPRD} a_1 < x {\oPRD} a \RW
          z_1^2 = a_1 \leq x_1 \RW
          f(z_1,a_1,x_1)^2 \leq x_1 < (f(z_1,a_1,x_1){+}1)^2)}{\ (IH)}
 }{
 \pstreeJKa{
  \TrJKl{}{z^2 = a \leq x}{\ \LABEL{eq:ex:mind:sqrt:note}}
  }{
  \pstreeJKa{
   \TrJKl{}{f(z,a,x)^2 \leq x < (f(z,a,x)+1)^2}
         {\GOAL\, \LABEL{eq:ex:mind:sqrt:spec:proof:goal}}
   }{
   \pstreeJKa{
    \TrJK{}{a+2 \oTIMES z+1 \SETr a_1}
    }{
    \pstreeJKb{
     \Tr{}
     }{
     \pstreeJKa{
      \TrJK{}{x < a_1}
      }{
      \pstreeJKai{
       \TrJK{}{f(z,a,x) \SETl z}
       }{
       }
      }
     \pstreeJKa{
      \TrJK{}{x \geq a_1}
      }{
      \pstreeJKa{
       \TrJKl{}{f(z+1,a_1,x)^2 \leq x < (f(z+1,a_1,x)+1)^2}
               {\ \LABEL{eq:ex:mind:sqrt:spec:proof:hyp}}
       }{
       \pstreeJKai{
        \TrJK{}{f(z,a,x) \SETl f(z+1,a_1,x)}
        }{
        }
       }
      }
     }
    }
   }
  }
 }
}
The proof assigns to $a_1$ so we have $a_1 = (z+1)^2$
and then does a dichotomy discrimination.
When $x < a_1$ the goal \XEL{ex:mind:sqrt}{spec:proof:goal}
holds trivially after defining $f(z,a,x) \SETl z$. 
When $x \geq a_1$ then $x \oPRD a_1 < x \oPRD a$
and we instantiate IH to \XEL{ex:mind:sqrt}{spec:proof:hyp} from which the
goal \XEL{ex:mind:sqrt}{spec:proof:goal} is obtained by the shown definition.
Note that the conclusion \XEL{ex:mind:sqrt}{note} of an assumption rule
is without computational effect because when $z^2 \neq a$ or $a > x$
holds then $f$ still terminates.

The square root function can be now explicitly derived by
$\aSQRT{x} = f(0,0,x)$. This is another example when a faster program can
be obtained by a detour through a partial function although
$f$ happens to be total in this case.

\PAR{An optimal program for \texorpdfstring{$\aSQRT{x}$}{[√x]} by \texorpdfstring{$4$}{4}-adic recursion}\label{par:ex:np}

All recursive definitions extracted so far share the same shortcoming: the
recursion goes exponentially longer than it should. Recursion on monadic
representation is computationally feasible only when initialized with
$\log$-sized arguments. Since we cannot restrict the square root function 
to small arguments, we have to use more economical recursion scheme.
Dyadic recursion will not work but
recursion in the base $4$ will because we have 
$\aSQRT{4\oTIMES x} \approx 2 \oTIMES \aSQRT{x}$. 
For a base $p > 1$ and an offset $m > 0$ we have  
the following rules
of induction on {\em $p$-adic notation\/}:
\begin{center}
\pstreeJKa{
 \TrJKl{}{\phi[x]}{\GOAL}
 }{
 \pstree[edge={\ncline[linestyle=dotted]},
         treesep=0.5cm, 
         treefit=loose, 
         levelsep=*0.2cm]{
  \Tr{}
  }{
  \pstreeJKai{
   \TrJKl{}{\phi[0]}{\GOAL}
   }{
   }
  \pstreeJKai{
   \TrJKl{}{\phi[1]}{\GOAL}
   }{
   }
  \pstreeJKai{
   \TrJK{edge=none}{\dotsb}
   }{
   }
  \pstreeJKai{
   \TrJKl{}{\phi[\NUM{m-1}]}{\GOAL}
   }{
   }
  \pstreeJKai{
   \TrJKl{}{\NUM{m} \leq i \wedge i < \NUM{m+p} \wedge
           \phi[x] \RW \phi[\NUM{p} \oTIMES x+i]}{\GOAL}
   }{
   }
  }
 }
\end{center}
We can extract the following fast definition of the square root function:
\begin{CLAUSES}
\> $\aSQRT{0} = 0$
\\ 
\> $\aSQRT{1} = 1$
\\ 
\> $\aSQRT{2} = 1$
\\ 
\> $\aSQRT{4 \oTIMES x+i} = z \LW 
  3 \leq i \wedge i < 7 \wedge 2 \oTIMES \aSQRT{x} = z \wedge 
  4 \oTIMES x+i < {(z+1)}^2$
\\ 
\> $\aSQRT{4 \oTIMES x+i} = z+1 \LW
  3 \leq i \wedge i < 7 \wedge 2 \oTIMES \aSQRT{x} = z \wedge 
  4 \oTIMES x+i \geq {(z+1)}^2\ \wedge \ $\\
\>\>\> $4 \oTIMES x+i < {(z+2)}^2$
\\ 
\> $\aSQRT{4 \oTIMES x+i} = z+2 \LW
  3 \leq i \wedge i < 7 \wedge 2 \oTIMES \aSQRT{x} = z \wedge 
  4 \oTIMES x+i \geq {(z+1)}^2\ \wedge \ $\\
\>\>\> $4 \oTIMES x+i \geq {(z+2)}^2$
\ .
\end{CLAUSES}
from a proof of its witnessing formula
by $p$-adic induction with $p=4$ and $m=3$.
We leave the base cases to the reader and show here only the inductive case:
\begin{center}\small
\pstreeJKa{
 \TrJK{}{3 \leq i < 7}
 }{
\pstreeJKa{
 \TrJKl{}{\aSQRT{x}^2 \leq x < (\aSQRT{x}+1)^2}{\ \ (IH)}
 }{
 \pstreeJKa{
  \TrJKl{}{\aSQRT{4 \oTIMES x + i}^2 \leq 4 \oTIMES x + i < 
       (\aSQRT{4 \oTIMES x + i}+1)^2}{\GOAL\, \LABEL{eq:ex:np:sqrt:proof}}
  }{
  \pstreeJKa{
   \TrJK{}{2 \oTIMES \aSQRT{x} \SETr z}
   }{
   \pstreeJKb{
    \Tr{}
    }{
    \pstreeJKa{
     \TrJK{}{4 \oTIMES x + i < (z+1)^2}
     }{
     \pstreeJKai{
       \TrJK{}{\aSQRT{4 \oTIMES x+i} \SETl z}
      }{
      }
     }
    \pstree[edge=none,
         treesep=1.5cm,
         treefit=loose,
         levelsep=*0.0cm
         ]{
     \TrJK{}{4 \oTIMES x + i \geq (z+1)^2}
     }{
     \pstreeJKb{
      \Tr{}
      }{
      \pstreeJKa{
       \TrJK{}{4 \oTIMES x + i < (z+2)^2}
       }{
       \pstreeJKai{
         \TrJK{}{\aSQRT{4 \oTIMES x+i} \SETl z+1}
        }{
        }
       }
      \pstreeJKa{
       \TrJK{}{4 \oTIMES x + i \geq (z+2)^2}
       }{
       \pstreeJKai{
         \TrJK{}{\aSQRT{4 \oTIMES x+i} \SETl z+2}
        }{
        }
       }
      }
     }
    }
   }
  }
 }
 }
\end{center}
After assignment $2 \oTIMES \aSQRT{x} \SETr z$ we do a dichotomy discrimination.
When
$4 \oTIMES x{+}i < (z{+}1)^2$
we define $\aSQRT{4 \oTIMES x{+}i} \SETl z$ and prove the
goal \XEL{ex:np}{sqrt:proof} by:
\[
\aSQRT{4 \oTIMES x + i}^2 =
z^2 =
4 \oTIMES \aSQRT{x}^2 \stackrel{IH}{\leq} 4 \oTIMES x <
4 \oTIMES x + i <
(z+1)^2 =
(\aSQRT{4 \oTIMES x + i}+1)^2
\ .
\]
When $4 \oTIMES x {+} i \geq (z{+}1)^2$ then we do another 
dichotomy discrimination.
When $4 \oTIMES x {+} i < (z{+}2)^2$ we define
$\aSQRT{4 \oTIMES x{+}i} \SETl z{+}1$ and prove the goal trivially.
Finally, when $4 \oTIMES x{+}i \geq (z{+}2)^2$
then we define $\aSQRT{4 \oTIMES x{+}i} \SETl z{+}2$ and prove the
the first half of the goal 
trivially.
For the second half we have
$x < (\aSQRT{x}{+}1)^2$ from IH and thus
$x \leq \aSQRT{x}^2{+}2 \oTIMES \aSQRT{x}$.
Therefore we get:
\begin{align*}
4 \oTIMES x + i & \leq 
4 \oTIMES \aSQRT{x}^2 + 8 \oTIMES \aSQRT{x} + i < 
4 \oTIMES \aSQRT{x}^2 + 12 \oTIMES \aSQRT{x} + 9 =
\\
& (2 \oTIMES \aSQRT{x} + 3)^2 =
(z+3)^2  =
(\aSQRT{4 \oTIMES x + i} + 1)^2
\ .
\end{align*}

\PAR{Induction on pair notation}\label{par:ex:p}

The rules of induction on {\em pair notation\/} permit to extraction
programs with computationally feasible recursion:
\MYPSa{-0.2}{-0.1}{
\pstreeJKa{
 \TrJKl{}{\phi[x]}{\GOAL}
 }{
 \pstreeJKb{
  \Tr{}
  }{
  \pstreeJKai{
   \TrJKl{}{\phi[0]}{\GOAL}
   }{
   }
  \pstreeJKai{
   \TrJKl{}{\phi[v] \wedge \phi[w] \RW \phi[(v,w)]}{\GOAL}
   }{
   }
  }
 }
}
For instance, the function $x \oplus y$ concatenating two lists is 
defined primitively recursively by {\em recursion on pair notation\/}:
\(
0 \oplus y = y
\)
and 
\(
(v,w) \oplus y = v,w \oplus w
\).
We can easily prove by pair induction properties of concatenation, such as
the associativity:
\(
x \oplus (y \oplus z) = (x \oplus y) \oplus z
\).  

Consider the well-known function $\nRT(x)$ {\em flattening\/} the number $x$
into a list of the pair size $\aSZP{x}$ and containing only zeroes as
elements. The function is defined primitively recursively by pair recursion:
\begin{CLAUSES}
\> $\nRT(0) = 0$
\\
\> $\nRT(v,w) = 0,\nRT(v) \oplus \nRT(w)$
\ .
\end{CLAUSES}
The program runs in time $O(\aSZP{x}^2)$ due to repeated concatenations
whereas $0(\aSZP{x})$ suffices when we extract the accumulator version
$f(x,a)$ of $\nRT$ from the following proof by pair recursion on $x$ with
the witnessing formula for the {\em $\Pi_2$-specification formula\/}
\(
\forall a \exists y\, y = \nRT(x) \oplus a
\): 
\MYPSa{-0.2}{-0.1}{
\pstreeJKa{
 \TrJKl{}{\forall a\, f(x,a) = \nRT(x) \oplus a}{\GOAL}
 }{
 \pstreeJKb{
  \Tr{}
  }{
  \pstreeJKa{
   \TrJKl{}{f(0,a) = \nRT(0) \oplus a}{\GOAL\ \LABEL{eq:ex:p:base}}
   }{
   \pstreeJKai{
    \TrJK{}{f(0,a) \SETl a}
    }{
    }
   }  
  \pstreeJKa{
   \TrJKl{}{\forall a\, f(v,a) = \nRT(v) \oplus a}{\ \ ($IH_1$)}
   }{
   \pstreeJKa{
    \TrJKl{}{\forall a\, f(w,a) = \nRT(w) \oplus a}{\ \ ($IH_2$)}
    }{
    \pstreeJKa{
     \TrJKl{}{f((v,w),a) = \nRT(v,w) \oplus a}{\GOAL\ \LABEL{eq:ex:p:ind}}
     }{
     \pstreeJKai{
      \TrJK{}{f((v,w),a) \SETl 0,f(v,f(w,a))}
      }{
      }
     }
    }
   }
  }
 }
}
Of course, we can extract $f$ explicitly by defining $f(x,a) \SETl \nRT(x) \oplus a$
but that runs in time $O(\aSZP{x}^2)$.
The goal \XEL{ex:p}{base} in the base case is trivially satisfied by defining
$f(0,a) \SETl a$. In the inductive case
we satisfy the goal \XEL{ex:p}{ind}  by the shown definition and 
by instantiating both induction hypotheses:
\begin{align*}
f((v,w),a) \stackrel{\text{def}}{=} {} & 0,f(v,f(w,a)) \stackrel{IH_2}{=} 
 0,f(v,\nRT(w) \oplus a) \stackrel{IH_1}{=} \\
 & 0,\nRT(v) \oplus \nRT(w) \oplus a  =
  \nRT(v,w) \oplus a  
\end{align*}
The reader will note that we have instantiated $IH_1$ with $a \SETl \nRT(w) \oplus a$
and so the induction formula must be $\Pi_2$.
The extracted program for $f$ is by {\em simply nested\/} 
recursion on pair notation and $\nRT$ is explicitly defined from $f$:
\begin{CLAUSES}
\> $f(0,a) = a$
\\
\> $f((v,w),a) = 0,f(v,f(w,a))$
\\
\> $\nRT(x) = f(x,0)$
\ .
\end{CLAUSES}

\PAR{Binary trees}\label{par:ex:bt}

We arithmetize binary trees with labels from $N$ as in
\cite{qq:KomaraVoda:98:CSL98} by two constructors:
$E = 0,0$ (the empty tree)
and
$\nND(x,l,r) = 1,x,l,r$ (a node with label $x$ and two subtrees $l$ and $r$).
The predicate \nBT\ holding of codes of binary trees is defined 
primitively recursively by {\em course of values recursion\/}:
\begin{CLAUSES}
\> $\nBT(E)$
\\
\> $\nBT\, \nND(x,l,r) \LW \nBT(l) \wedge \nBT(r)$
\ .
\end{CLAUSES}
This is a definition of an {\em inductive\/} predicate which affords
rules of {\em \nBT-induction\/}:
\MYPSa{-0.4}{-0.2}{
\pstreeJKa{
 \TrJKl{}{\nBT(t) \RW \phi[t]}{\GOAL}
 }{
 \pstreeJKb{
  \Tr{}
  }{
  \pstreeJKai{
   \TrJKl{}{\phi[E]}{\GOAL}
   }{
   }
  \pstreeJKai{
   \TrJKl{}{\phi[l] \wedge \phi[r] \RW \phi[\nND(x,l,r)]}{\GOAL}
   }{
   }
  }
 }
}
The function \aSZBT{t} counting the number of labels and the predicate $x
\rINBT t $ of binary tree membership are defined primitively recursively by
{\em \nBT-recursion\/}:
\begin{CLAUSES}
\> $\aSZBT{E} = 0$
\>\>\> $x \rINBT \nND(y,l,r) \LW
          x = y$
\\
\> $\aSZBT{\nND(x,l,r)} = \aSZBT{l}+\aSZBT{r}+1$
\>\>\> $x \rINBT \nND(y,l,r) \LW
          x \neq y \wedge x \rINBT l$
\\
\>
\>\>\> $x \rINBT \nND(y,l,r) \LW
          x \neq y \wedge x \not\rINBT l \wedge x \rINBT r$
\ .
\end{CLAUSES}
The close relationship between the programming constructs and our tableau
proof rules is nicely demonstrated by the proof of the `boundedness' property
of binary trees:
\(
x \rINBT t \RW x < t
\)
which is by complete induction on $t$ and uses the entire clausal definition
of $x \rINBT t$ for discrimination because under that assumption exactly one
of the clauses applies:
\MYPSa{-0.4}{-0.2}{
\pstreeJKa{
 \TrJKl{}{\forall t_1(t_1 < t \RW x \rINBT t_1 \RW x < t_1)}{
          \ (IH)}
 }{
 \pstreeJKa{
  \TrJK{}{x \rINBT t}
  }{
  \pstreeJKa{
   \TrJKl{}{x < t}{\GOAL}
   }{
   \pstree[edge={\ncline[linestyle=dotted]},
           treesep=1.5cm,
           treefit=loose,
           levelsep=*0.2cm  
           ]{
    \Tr{}
    }{
    \pstreeJKa{
     \TrJK{}{t = \nND(y,l,r)}
     }{
     \pstreeJKai{
     \TrJK{}{x = y}
      }{
      }
     }
    \pstreeJKa{
     \TrJK{}{t = \nND(y,l,r)}
     }{
     \pstreeJKa{
      \TrJK{}{x \neq  y \wedge x \rINBT l}
      }{
      \pstreeJKa{
       \TrJK{}{[t_1 \SETl l \text{\ in $(IH)$}]}
       }{
       }
      }
     }
    \pstreeJKa{
     \TrJK{}{t = \nND(y,l,r)}
     }{
     \pstreeJKa{
      \TrJK{}{x \neq  y \wedge x \not\rINBT l \wedge x \rINBT r}
      }{
      \pstreeJKai{
       \TrJK{}{[t_1 \SETl r \text{\ in $(IH)$}]}
       }{
       }
      }
     }
    }
   }
  }
 }
}
The following explicitly defined predicates are primitive recursive
because the quantifiers are {\em bounded\/}:
\begin{equation*}
\aLTBST{t}{x} \LRW \forall y \rINBT t\,.\ y < x
\text{\makebox[1.0cm]{}}
\aGTBST{t}{x} \LRW \forall y \rINBT t\,.\ y > x
\ .
\end{equation*}

\PAR{Binary search trees}\label{par:ex:bst}

We arithmetize binary search trees by codes $t$ satisfying the 
predicate $\nBST(t)$ primitively recursively defined by \nBT-recursion:
\begin{CLAUSES}
\> $\nBST(E)$
\\
\> $\nBST\, \nND(x,l,r) \LW
  \aLTBST{l}{x} \wedge \aGTBST{r}{x} \wedge \nBST(l) \wedge \nBST(r)$
\ .
\end{CLAUSES}
$\nBST$ is an inductive predicate affording the following
rules of {\em \nBST-induction\/}:
\MYPSa{-0.4}{-0.2}{
\pstreeJKa{
 \TrJKl{}{\nBST(t) \RW \phi[t]}{\GOAL}
 }{
 \pstreeJKb{
  \Tr{}
  }{
  \pstreeJKai{
   \TrJKl{}{\phi[E]}{\GOAL}
   }{
   }
  \pstreeJKai{
   \TrJKl{}{\aLTBST{l}{x} \wedge \aGTBST{r}{x} \wedge \phi[l] \wedge
           \phi[r] \RW\phi[\nND(x,l,r)]}{\GOAL}
   }{
   }
  }
 }
}
By straightforward \nBST-induction we can prove $\nBST(t) \RW \nBT(t)$ asserting
that binary search trees form a {\em subsort\/} of binary trees.

We can extract from \ISA-proofs also definitions of predicates
because they are introduced through their characteristic functions ($0$ is false,
$1$ is true). 
For instance, the program for the predicate $x \rINBST t$ 
extracted as an `unknown'
predicate from the following proof 
can speed up the
$O(\aSZBT{t})$ search of $x \rINBT t$ up to
$O(\log \aSZBT{t})$:
\MYPS{-0.1}{-0.1}{
\pstree[edge={\ncline[linestyle=dotted]},
        treesep=-2.0cm, 
        treefit=loose, 
        levelsep=*0.2cm
        ]{
 \TrJKl{}{\nBST(t) \RW (x \rINBST t \LRW  x \rINBT t)}{\GOAL}
 }{
 \pstreeJKa{
  \TrJKl{}{x \rINBST E \LRW  x \rINBT E}{\GOAL}
  }{
  \TrJK{}{x \rINBST E \SETl \bot}
  }
 \pstreeJKa{
  \TrJK{}{\aLTBST{l}{y} \wedge \aGTBST{r}{y}}
  }{
  \pstreeJKa{
   \TrJKl{}{x \rINBST l \LRW  x \rINBT l}{\ \LABEL{eq:ex:bst:inbst:wtn:ih:l}}
   }{
    \pstreeJKa{ 
    \TrJK{}{x \rINBST r \LRW  x \rINBT r}
    }{
    \pstreeJKa{
     \TrJKl{}{x \rINBST \nND(y,l,r) \LRW  x \rINBT \nND(y,l,r)}{
              \GOAL\, \LABEL{eq:ex:bst:inbst:wtn:ic}}
     }{
     \pstree[edge={\ncline[linestyle=dotted]},
             treesep=0.4cm,
             treefit=loose,
             levelsep=*0.2cm
             ]{
      \Tr{}
      }{
      \pstreeJKa{
       \TrJK{}{x < y}
       }{
       \pstreeJKa{
        \TrJK{}{x \rINBST \nND(y,l,r) \SETl x \rINBST l}
        }{
         }
        }
      \pstreeJKa{
       \TrJK{}{x = y}
       }{
       \pstreeJKai{
        \TrJK{}{x \rINBST \nND(y,l,r) \SETl \top}
        }{
        }
       }
      \pstreeJKa{
       \TrJK{}{x > y}
       }{ 
        \pstreeJKai{
         \TrJK{}{x \rINBST \nND(y,l,r) \SETl x \rINBST r}
        }{ 
        }
       }
      }
     }
    }
   } 
  }
 }
}
The proof is by \nBST-induction followed by a trichotomy discrimination.
Note the definitions of the unknown truth values of $x \rINBST t$.
For instance, the goal \XEL{ex:bst}{inbst:wtn:ic} is
proved in the case $x < y$ by:
\begin{align*}
x \rINBST \nND(y,l,r) \EQx{def}{\mLRW}
x \rINBST l  \EQx{IH:\XEL{ex:bst}{inbst:wtn:ih:l}}{\mLRW}
x \rINBT l \EQx{$\aGTBST{r}{y}$}{\mLRW}
x \rINBT \nND(y,l,r) 
\ .
\end{align*}
The extracted definition with the negative clause omitted by default is:
\begin{CLAUSES}
\> $x \rINBST \nND(y,l,r) \LW  x < y \wedge x \rINBST l$
\\
\> $x \rINBST \nND(y,l,r) \LW  x = y$
\\
\> $x \rINBST \nND(y,l,r) \LW  x > y \wedge x \rINBST r$
\ .
\end{CLAUSES}

The function \aINS{x}{t} inserting $x$
into the binary search tree $t$
satisfies:
\[ \nBST(t) \RW \nBST(\aINS{x}{t})  \wedge 
      \forall z ( z \rINBST \aINS{x}{t} \LRW z = x \vee z \rINBST t)\ .
\]
The quantifier $\forall z$ is bounded and so we can extract from
a proof by \nBST-induction on $t$ the following program:
\begin{CLAUSES}
\> $\aINS{x}{E} = \nND(x,E,E)$
\\
\> $\aINS{x}{\nND(y,l,r)} = \nND(y,\aINS{x}{l},r) \LW x < y$
\\
\> $\aINS{x}{\nND(y,l,r)} = \nND(y,l,r) \LW x = y$
\\
\> $\aINS{x}{\nND(y,l,r)} = \nND(y,l,\aINS{x}{r}) \LW x > y$
\ .
\end{CLAUSES}

%% file: sc4.tex
\SECT{Admissibility of Extraction Rules and of Induction Rules in 
\texorpdfstring{\ISA}{IΣ₁}-arithmetic}\label{sc:char}

Our tableau proof system uses
extraction
rules
and
a rich set of 
induction rules and we now briefly sketch 
their admissibility in \ISA-arithmetic.

\PAR{Admissibility of extraction rules}
Extraction proofs \XEG{specs}{witness} are translated to 
the standard \ISA-proofs \XEG{specs}{trans} as outlined in \XP{specs}.
Applications of potentially partial function symbols $f$ in the
terms of extraction proofs are only a shorthand to improve the 
readability and are eliminable by the graph of $f$.
Induction hypotheses $\phi[\ARR{\rho},f(\ARR{\rho})]$ are thus
of the form $\exists y(\phi[\ARR{\rho},y] \wedge f(\ARR{\rho}) \GR y)$ 
asserting that $f(\ARR{\rho})$ is defined.  

\PAR{Admissibility of basic induction rules}\label{par:char:ind}

Our proof system is tableau-based and so it has induction rules with side
formulas rather than the induction axioms of \ISA-arithmetic. It should be
obvious that the induction formula $\phi[x]$ of the rule of mathematical
induction in \XP{ex:n} can be any $\Sigma_1$-formula with arbitrary side
formulas.  $\Pi_1$-induction rules with arbitrary side formulas reduce to
$\Sigma_1$-rules the same way as axioms do (see \cite{qq:HajekPudlak:93}).
Weak $\Pi_2$-induction rules as defined in \cite{qq:KomaraVoda:98:CSL98} are
admissible; their side formulas
(including nested induction)
must be weak, i.e. $\Pi_1$ in assumptions and
$\Sigma_1$ in goals. The same restrictions apply to complete, dyadic, and
pair inductions because the first reduces to mathematical induction and the
last two reduce directly to complete induction. $\Pi_2$-induction rules are
necessary when the extracted programs have
substitution in parameters or
nested recursion
(see \XP{ex:p}).

\PAR{Admissibility of rules of measure induction}\label{par:char:m}

A rule of measure induction (see \XP{ex:mind}) 
with the induction formula $\phi[\ARR{x}]$ reduces
to the rule of mathematical induction on $n$ with the formula:
$\forall \ARR{x} (m(\ARR{x}) < n \RW \phi[\ARR{x}])$.
Hence, if $\phi$ is $\Sigma_1$ or $\Pi_2$ the side formulas must be weak.

\PAR{Admissibility of \texorpdfstring{$R$}{R}-induction rules}\label{par:char:r}

After the elimination of
variables introduced into the clauses by assignments and by discrimination on
patterns
every clausal definition of a primitive recursive predicate $R$ can be
brought into the form
\vspace{-1mm}
{\small
\begin{equation}\label{eq:char:r:def}
R(\ARR{x}) \LRW \bigvee_{i=1}^k \psi_i[R(\cdot)\,;\ARR{x}]
\end{equation}
} \!\!\!\!\!
with $\psi_i$ conjunctions of literals.
For instance, the pattern $\tau = x+1$ is replaced by $\tau > 0$
and the variable $x$ is eliminated by $\tau \oPRD 1$.

The predicate $R$ is {\em inductive\/} if all recursive applications in
\(
\bigvee_{i=1}^k \psi_i
\)
are positive. The inductive predicate $R$ affords the rules
of {\em $R$-induction\/}:
\vspace{2mm}
\\
{\small
\begin{minipage}[c]{140mm}
\begin{minipage}[c]{50mm}
\begin{center}
\pstreeJKa{
 \TrJKl{}{R(\ARR{x}) \RW \phi[\ARR{x}]}{\GOAL}
 }{
 \pstree[edge={\ncline[linestyle=dotted]},
         treesep=0.2cm,
         treefit=loose,
         levelsep=*0.3cm
        ]{
  \Tr{}
  }{
  \pstree[edge=none,
          levelsep=*0.2cm,
          treesep=0.3cm]{
   \TrJKl{}{\bigvee_{i=1}^k \psi_i[\phi[\cdot];\ARR{x}] \RW \phi[\ARR{x}]}{\GOAL}
   }{
    \pstree[edge=none,
            levelsep=*0.2cm,
            treesep=0.3cm]{
     \Ttri[trimode=U]{$p$}
     }{
     }

   }
  }
 }
\end{center}
\end{minipage}
\begin{minipage}[c]{60mm}
$\phi$ is $\Sigma_1$, $\Pi_1$, or $\Pi_2$,
where if $\phi$ is $\Pi_2$ then the side formulas must be weak.
\end{minipage}
\begin{minipage}[c]{20mm}
\ \ \ \ \ \!
\LABEL{eq:char:r:ind}
\end{minipage}
\end{minipage}
}
\\
Boyer and Moore \cite{qq:BoyerMoore:79} were the first to use inductive
rules derived from recursive definitions. Similar rules
are used in HOL \cite{qq:GordonMelhalm:93:full}.

The soundness of the rule \XEL{char}{r:ind} follows from the fact 
that $R$ is the minimal predicate satisfying
the $(\LW)$-implication of \XEL{char}{r:def}. 
This is sufficient for the admissibility of \XEL{char}{r:ind}
in the system \cite{qq:BoyerMoore:79}. 
The admissibility is proved in HOL
in a rather
strong Simple Theory of Types (whose
strength is similar to ZF).
Our task is substantially harder
because we have to prove the admissibility  
in a very weak theory \ISA.
For that we use an auxiliary
primitive recursive
predicate $\bar R(i,\ARR{x})$
approximating $R$ and satisfying:
{\small
\begin{equation*}\label{eq:char:r:ar:def}
\neg \bar R(0,\ARR{x})
\quad
\text{and}
\quad
\bar R(i+1,\ARR{x}) \LRW 
\bigvee_{i=1}^k \psi_i[\bar R(i,\cdot);\ARR{x}] \ .
\end{equation*}
} \!\!\!\!\!
\ISA\ clearly proves 
\(
R(\ARR{x}) \LRW \exists i\,\bar R(i,\ARR{x})
\)
and so the admissibility of \XEL{char:r}{ind}
reduces to the problem of proving in \ISA\ 
\begin{equation}\label{eq:char:r:ar:thm}
\forall \ARR{x} (\bar R(i,\ARR{x}) \RW \phi[\ARR{x}]) 
\end{equation}
from the proof $p$ of
\(
\bigvee_{i=1}^k \psi_i[\phi[\cdot];\ARR{x}] \RW \phi[\ARR{x}]
\).
This is done by bounding in \ISA\  the quantifiers $\forall \ARR{x}$ of
\XEL{char:r}{ar:thm} with the help of a primitive
recursive {\em course of values\/} function $f(i,\ARR{z})$
which yields a list containing the pair $(a_1,\ldots,a_n)$ for all 
computational `predecessors' $\ARR{a}$ of
$\ARR{z}$ at the $i$-th level of recursion. The function satisfies:
\begin{equation*}\label{eq:char:r:f:def}
(\ARR{x}) \IN f(0,\ARR{z}) \LRW
(\ARR{x}) = (\ARR{z})
\quad
\text{and}
\quad
(\ARR{x}) \IN f(i,\ARR{z}) \RW (\ARR{\rho}[\ARR{x}]) \IN f(i+1,\ARR{z})
\end{equation*}
for all recursive applications $R(\ARR{\rho}[\ARR{x}])$ in
$\bigvee_{i=1}^k \psi_i$.
Here $x \IN y$ is the list membership predicate.
\ISA\ then proves by induction on $i$:
\begin{equation*}\label{eq:char:r:ar:lemma}
\forall j\leq m\, \forall x_1{\leq} f(j,\ARR{z})\ldots
\forall x_n{\leq} f(j,\ARR{z})\,\bigl(
(\ARR{x}) \IN f(j,\ARR{z}) \wedge 
i+j = m \wedge \bar R(i,\ARR{x}) \RW \phi[\ARR{x}]\,\bigr)
\ .
\end{equation*}
Now \XEL{char:r}{ar:thm} follows in \ISA\ 
by setting
$m:=i$,
$\ARR{z}:=\ARR{x}$,
and
$j:=0$.
Note that we have actually proved that $R$-induction is admissible
(with weak side formulas if $\phi$ is $\Sigma_1$)
even if
$R$ is a recursively enumerable predicate although such an $R$ 
cannot be used in clausal definitions.

\SECT{Conclusion}
We have hopefully demonstrated that the seemingly weak
$\PR+\ISA$ paradigm is
sufficient to introduce a usable programming language and its verification theory.
We have implemented the paradigm in the system CL which we use in
four undergraduate
courses at our university. The courses are: Introduction to declarative programming,
Predicate Logic, Program specification and verification, and Theory of
computability. They are taken every year by about 250 students.
The main reason why we can teach this is that the intuition
about the true properties
of CL programs is easily acquired by the students. They do not need
to know more than the
standard model of $\N$.

We hope that we have convinced the reader that we can arithmetize in $\N$ the
basic data structures needed in the computer programming with the level of
comfort comparable to that in the declarative programming languages.
The objection that our system proves the termination of only primitive recursive
functions is answered by pointing out that the computationally feasible functions
are a very small subset of \PR\ functions. Besides, by the Incompleteness theorem of
G\"odel, no formal theory can prove the termination of all recursive functions. 

We have also built into CL a
mechanism for the definition of abstract data types (ADT) through
non-recursive extensions of \ISA-arithmetic (this defines ADT's). The
consistency of extensions is proved by primitive recursive interpretations
into \ISA\ (this constitutes implementations of ADT's). 

In the close future we plan to implement the automatic extraction
of witnessing functions from $\Pi_2$-specifications as discussed in this paper.
This can be
done without the loss of efficiency when compared with hand-coded programs
but with the added advantage that we prove at the same
time the total correctness of the function being constructed. We justify our
approach of leaving to the user the decision about the computational content
of rules by the current trend in the design of theorem provers. Namely,
the first theorem
provers, for instance the system of Boyer and Moore, tried to construct the
proofs automatically. This has proved to be untenable and so the newer
systems such as NuPrl, PX, PVS, Coq, Isabelle, MINLOG, HOL, and CL are
intelligent proof checkers where the user guides the proofs.

We further plan to add
to CL the intensional functionals (via coding into $\N$ in the style of lambdas
of LISP).
We will then have demonstrated that CL can
do most of the things declarative languages can do but with the simple
semantics of a rather weak formal theory.
We are currently preparing an extension of CL to deal
with typed programs and
we plan to compile the CL programs with the 
in-place-modification of data structures.

%% file: ms.bbl
\newcommand{\etalchar}[1]{$^{#1}$}
\newcommand{\noopsort}[1]{} \newcommand{\singletter}[1]{#1}

%% file: ms.bbl
\begin{thebibliography}{BBS{\etalchar{+}}98}

\bibitem[BBS{\etalchar{+}}98]{qq:MINLOG:98}
H.~Benl, U.~Berger, H.~Schwichtenberg, M.~Seisenberger, and W.~Zuber.
\newblock Proof theory at work: {P}rogram development in the {MINLOG} system.
\newblock In W.~Bibel and P.H. Schmitt, editors, {\em Automated Deduction},
  volume~II. Kluwer, 1998.

\bibitem[BM79]{qq:BoyerMoore:79}
R.~S. Boyer and J.~S. Moore.
\newblock {\em A Computational Logic}.
\newblock Academic Press, 1979.

\bibitem[GM93]{qq:GordonMelhalm:93:full}
M.~J.~C. Gordon and T.~F. Melhalm, editors.
\newblock {\em Introduction to {HOL}: {A} {T}heorem {P}roving {E}nvironment for
  {H}igher-{O}rder {L}ogic}.
\newblock Cambridge University Press, 1993.

\bibitem[HN88]{qq:PX:88}
S.~Hayashi and H.~Nakano.
\newblock {\em {PX}: A Computational Logic}.
\newblock The MIT Press, 1988.

\bibitem[HP93]{qq:HajekPudlak:93}
P.~H\'{a}jek and P.~Pudl\'{a}k.
\newblock {\em Metamathematics of First-Order Arithmetic}.
\newblock Springer Verlag, 1993.

\bibitem[Kre52]{qq:Kreisel:52}
G.~Kreisel.
\newblock On the interpretation of non-finitist proofs {II}.
\newblock {\em Journal of Symbolic Logic}, 17:43--58, 1952.

\bibitem[KV99]{qq:KomaraVoda:98:CSL98}
J.~Komara and P.~J. Voda.
\newblock Theorems of {P\'eter} and {Parsons} in computer programming.
\newblock In G.~Gottlob, E.~Grandjean, and K.~Seyr, editors, {\em Proceedings
  of CSL'98}, number 1584 in LNCS, pages 204--223. Springer Verlag, 1999.

\bibitem[P{\'e}t67]{qq:Peter:67}
R.~P{\'e}ter.
\newblock {\em Recursive Functions}.
\newblock Academic Press, 1967.

\bibitem[Ros82]{qq:Rose:84}
H.~E. Rose.
\newblock {\em Subrecursion: Functions and Hierarchies}.
\newblock Number~9 in Oxford Logic Guides. Clarendon Press, Oxford, 1982.

\bibitem[Sho67]{qq:Shoenfield:67}
J.~R. Shoenfield.
\newblock {\em Mathematical Logic}.
\newblock Addison-Wesley, 1967.

\bibitem[Smu68]{qq:Smullyan:68}
R.~Smullyan.
\newblock {\em First Order Logic}.
\newblock Springer Verlag, 1968.

\bibitem[Vod95]{qq:Voda:94:CSL94}
P.~J. Voda.
\newblock Subrecursion as a basis for a feasible programming language.
\newblock In L.~Pacholski and J.~Tiuryn, editors, {\em Proceedings of CSL'94},
  number 933 in LNCS, pages 324--338. Springer Verlag, 1995.

\bibitem[Vod00]{qq:Voda:theory:00:www}
P.~J. Voda.
\newblock Theory of {R}ecursive {F}unctions \& {C}omputability (from {C}omputer
  {P}rogrammer's {V}iew), 2000.
\newblock Available through WWW from
  {http://dent.ii.fmph.uniba.sk/$\thicksim$voda/text.ps.gz}.

\end{thebibliography}
